\documentclass{epl2}

\def\av#1{\left\langle#1\right\rangle}
\usepackage{latexsym}
\usepackage{amstext,amssymb,amsfonts}
\usepackage{epsfig}

\DeclareRobustCommand{\baselinestretch{2.2}}

\title{Transport in networks with multiple sources and sinks}

\author{Shai Carmi\inst{1,2} \and Zhenhua Wu\inst{2} \and Shlomo
 Havlin\inst{1} \and H. Eugene Stanley\inst{2}}

\shortauthor{S. Carmi \etal}

\institute{
 \inst{1}Minerva Center \& Department of Physics, Bar-Ilan
University, Ramat Gan 52900, Israel\\
 \inst{2}Center for Polymer
Studies \& Department of Physics, Boston University, Boston, MA
02215 USA}


\abstract{We investigate the electrical current and flow (number of
parallel paths) between two sets of $n$ sources and $n$ sinks in
complex networks. We derive analytical formulas for the average
current and flow as a function of $n$. We show that for small $n$,
increasing $n$ improves the total transport in the network, while
for large $n$ bottlenecks begin to form. For the case of flow, this
leads to an optimal $n^*$ above which the transport is less
efficient. For current, the typical decrease in the length of the
connecting paths for large $n$ compensates for the effect of the
bottlenecks. We also derive an expression for the average flow as a
function of $n$ under the common limitation that transport takes
place between specific pairs of sources and sinks.}

\date{\today}

\begin{document}

\maketitle

Transport processes, such as electrical current, diffusion, and
flow, are fundamental in physics, chemistry, and biology. Transport
properties depend critically on the structure of the medium, and
have been studied for a large variety of geometries \cite{DbA_Book}.
Of current interest are situations where the transport occurs on a
network. For example, information is transferred over computer or
social networks, vehicles traverse transportation networks, and
electrical current flows in power-grid networks.  Therefore,
understanding of mechanisms to increase transport effectiveness is
of great importance.

Transport properties usually have been investigated in the context
of transport between a single pair comprising one source and one
sink
\cite{klafter_nature,Lopez,Li,lazaros_pnas,Dani,Rieger_rw,Rieger}.
The quality of transport strongly depends on the degree (number of
connections) of the source and the sink, whereas the rest of the
network serves as an approximately resistance-free substrate for the
transport process. Consequently, the transport was found to strongly
depend on the degree distribution.

In more realistic situations, transport takes place between many
nodes simultaneously. For example, in peer-to-peer and other
computer networks users exchange files in parallel over the network
links. In transportation networks, vehicles travel between many
sources and many destinations through the network infrastructure.
The presence of many parallel transport processes on the same
underlying network leads to interactions between the different
deliveries and a change in network efficiency.  In this article we
will quantify this phenomenon analytically and numerically, and show
how different network usage leads to different behaviors. We
reported some preliminary results in \cite{My_flow2}.

We focus on the class of non-directed, non-weighted model networks
and we also study a real network. The first model is the
Erd\H{o}s-R\'{e}nyi (ER) network, in which each link exists with
independent small probability $p$. This leads to a Poisson degree
distribution \cite{ER,Bollobas}. The second model is the scale-free
(SF) network, characterized by a broad, power-law degree
distribution and was recently found to describe many natural systems
\cite{BA_review,DM_book,PV_book}. The ensemble of SF networks we
treat is the ``configuration model,'' in which node degrees are
drawn from a power-law distribution (see below), and then open links
are connected \cite{Molloy_Reed}. We also compute the distribution
of flows in a
 real network, the internet \cite{Medusa}.

We consider a transport process between two randomly chosen,
non-overlapping sets (sources and sinks) of nodes of size $n$ each,
where $1\leq n \leq N/2$, $N$ is the total number of nodes. We focus
on three explicit forms of transport, described below.

\emph{(i) Maximum flow} (henceforth denoted flow) between the
sources and sinks, when each link has unit capacity
\cite{Coreman,MF_algorithm,Rieger,Wu,My_flow2,Medusa}. For
non-weighted networks the flow is equivalent to the total number of
disjoint paths (i.e. paths that do not share any edge) that connect
the sources and sinks. Therefore, it quantifies any flow which does
not deteriorate with distance, such as flow of frictionless fluids,
traffic flow and information flow in communication networks.

\emph{(ii) Electrical current} in the network when the sources are
short circuited into a unit electrical potential, and the sinks to
the ground (assuming each link is a unit resistor). Electrical
current has special significance since it also describes any general
transport process in which the transport efficiency decreases with
the length of the path (due to increase of resistance). In addition,
the current is equal to the probability of a random walker starting
at any of the sources to escape to any of the sinks
\cite{Redner_book}.

\emph{(iii) Maximum multi-commodity (MC) flow}, where the sources
and sinks form ordered pairs, so the flow is directed from a given
source to a specific sink, and not to any other sink
\cite{Coreman,MCFlow_def} (here the network is directed and the
sources and sinks may overlap). This describes direct communication
between users in computer networks, or traffic of supplies over road
networks.

Our goal is to study the dependence of the total transport on the
number of sources/sinks in the three transport forms. We denote the
flow, electrical current, and MC flow by $F$, $I$, and
$F^{\textrm{MC}}$, respectively. In the case of a single pair, it
was shown \cite{Lopez,My_flow2} that the transport between a source
and a sink with degrees $k_1$ and $k_2$ is approximately:
$F_{n=1}(k_1,k_2) \approx F^{\textrm{MC}}_{n=1}(k_1,k_2) \approx
\min\{k_1,k_2\}$ for dense enough networks, and $I_{n=1}(k_1,k_2)
\approx c\frac{k_1k_2}{k_1+k_2}$, where $c \lesssim 1$ is a fitting
parameter. These equations state that the transport is dominated by
the degrees of the involved nodes, and the rest of the network is
practically a perfect conductor. Using the degree distribution:
$P(k) = e^{-\av{k}}\av{k}^k/k!$ for ER networks with average degree
$\av{k}$, and $P(k)\sim k^{-\gamma}~,~k\geq m$ for SF networks with
degree exponent $\gamma$ and minimum degree $m$, we can find the
distribution of flow or current.

Do transport properties change in the case of more than one pair?
For a small number of sources/sinks $n$, we expect no significant
difference. The backbone of the network remains an almost perfect
conductor, but the degree $k_1$ has to be replaced with the total
number of links emanating from the sources, and similarly for the
sinks \cite{My_flow2}. Because it is assumed that $n$ is small, we
neglect the possibility of internal links inside each set. Define
the \emph{sum} of $n$ degrees $z\equiv \sum_{i=1}^n{k_i}$. For ER
networks, $P_{Z}(z)= e^{-n\av{k}}(n\av{k})^z/z!$, a Poisson with
mean $n\av{k}$. For SF networks with $2<\gamma<3$, write $P(s)$ for
$s\rightarrow 1$, the generating function of $P(k)$ for large $k$,
as $P(s) \sim 1+A(1-s)+B(1-s)^{\gamma-1}+{\cal{O}}
\left((1-s)^{2}\right)$. Raising $P(s)$ to the power of $n$ yields
$P_{Z}(s) \sim 1+nA(1-s)+nB(1-s)^{\gamma -1}+{\cal{O}}
\left((1-s)^{2}\right)$, and thus for large $z$
\begin{equation}
\label{SFsum} P_{Z}(z)\sim z^{-\gamma}~,~z\geq nm.
\end{equation}

We define $z_1$ and $z_2$ to be the sum of degrees of the $n$ nodes
in the sources and sinks, respectively. Using these definitions, $F
= \min\{z_1,z_2\}$, and $I = c\frac{z_1z_2}{z_1+z_2}$. For the case
of MC flow, the pairs are independent and the total flow is the sum
of the flow of $n$ independent pairs.

The pdfs corresponding are therefore
\begin{eqnarray}
\label{transport_equation}
\label{flow_dist} \Phi_n(F) &=& 2\left[P_{Z}(F)\sum_{j\geq F}P_{Z}(j)\right] - [P_Z(F)]^2\\
\Phi_n(I) &=&
\sum_{z_1}\sum_{z_2}P_{Z}(z_1)P_{Z}(z_2)\delta(I-c\frac{z_1z_2}{z_1+z_2})
\\ \label{flow_dist_MC}
\Phi_n(F^{\textrm{MC}}) &=&
\mbox{Pr}\left\{\left[\sum_{j=1}^{n}\min\{k_{j,1},k_{j,2}\}\right] =
F^{\textrm{MC}}\right\},
\end{eqnarray}
where in Eq. (\ref{flow_dist_MC}) $k_{j,1}$ and $k_{j,2}$ are the
degrees of $j$th source and sink, respectively.

\begin{figure}[h]
\begin{center}
\includegraphics[width=5cm]{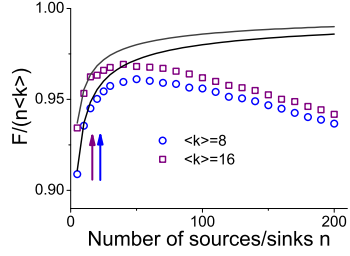} \vspace{-0.5cm}
\end{center}
\caption{Optimal number of sources/sinks. Average flow per
source/sink $\overline{F}/(n\av{k})$ vs. the number of sources/sinks
$n$. Symbols represent simulation results ($N=4096$, $\av{k}=8,16$;
average is taken over many realizations of the network and many
randomly chosen pairs), while lines represent the theory based on
the \emph{small-$n$ assumption} (Eq. (\ref{min_ER})). For $n
\lesssim \sqrt{N/\av{k}}$ (indicated with an arrow), the flow per
source/sink always increases, as predicted by the theory. However,
there is an optimal point beyond which the flow decreases.}
\label{Smalln}
\end{figure}

For ER networks we obtain a closed form formula for the flow pdf
\cite{My_flow2} {\small
\begin{equation}
\label{min_ER} \Phi_n(F) = 2 \frac{(n \av{k})^{F} e^{-n \av{k}}}{F!}
\left(\frac{\gamma(F,n\av{k})}{\Gamma(F)} - \frac{(n \av{k})^{F}
e^{-n \av{k}}}{2F!} \right),
\end{equation}}
where $\gamma(a,x)$ and $\Gamma(a)$ are the lower incomplete and
complete gamma functions, respectively. An interesting quantity to
study in a real transport system is the total average flow per
source/sink $\overline{F}(n)/n=\left[\sum_F F\Phi_n(F)\right]/n$,
since it represents the overall transport efficiency of the system.
For ER networks we calculate $\overline{F}(n)/n$ using Eq.
(\ref{min_ER}), and the results are compared with simulations in
Fig. \ref{Smalln}. The theoretical prediction is in good agreement
with the simulations for small values of $n$. For larger $n$, the
transport is less efficient than predicted, since interactions
between the paths begin to appear. Some paths become bottlenecks
\cite{bottleneck} and cannot serve more than their capacity.
Moreover, some links are wasted as they connect nodes within the
same set. In ER networks, the probability of having no intra-set
links is $(1-n/N)^{n\av{k}} \approx \exp[-n^2\av{k}/N]$. Therefore
already when $n \approx (N/\av{k})^{1/2}$, we expect some
deviations, as is confirmed by our simulations shown in Fig.
\ref{Smalln}.

\begin{figure}[h]
\begin{center}
\includegraphics[width=4cm]{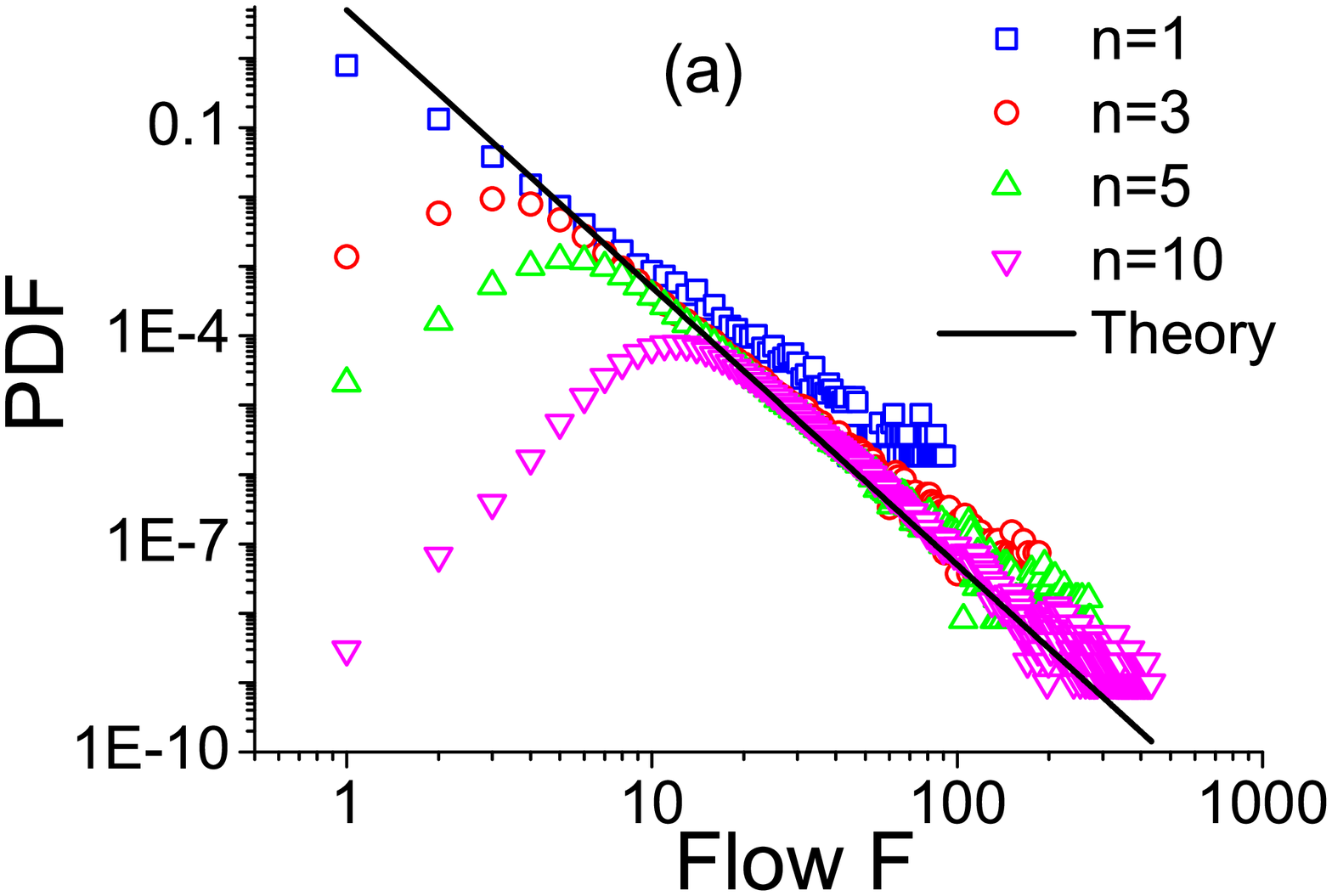}\vspace{-0.5cm}
\includegraphics[width=4cm]{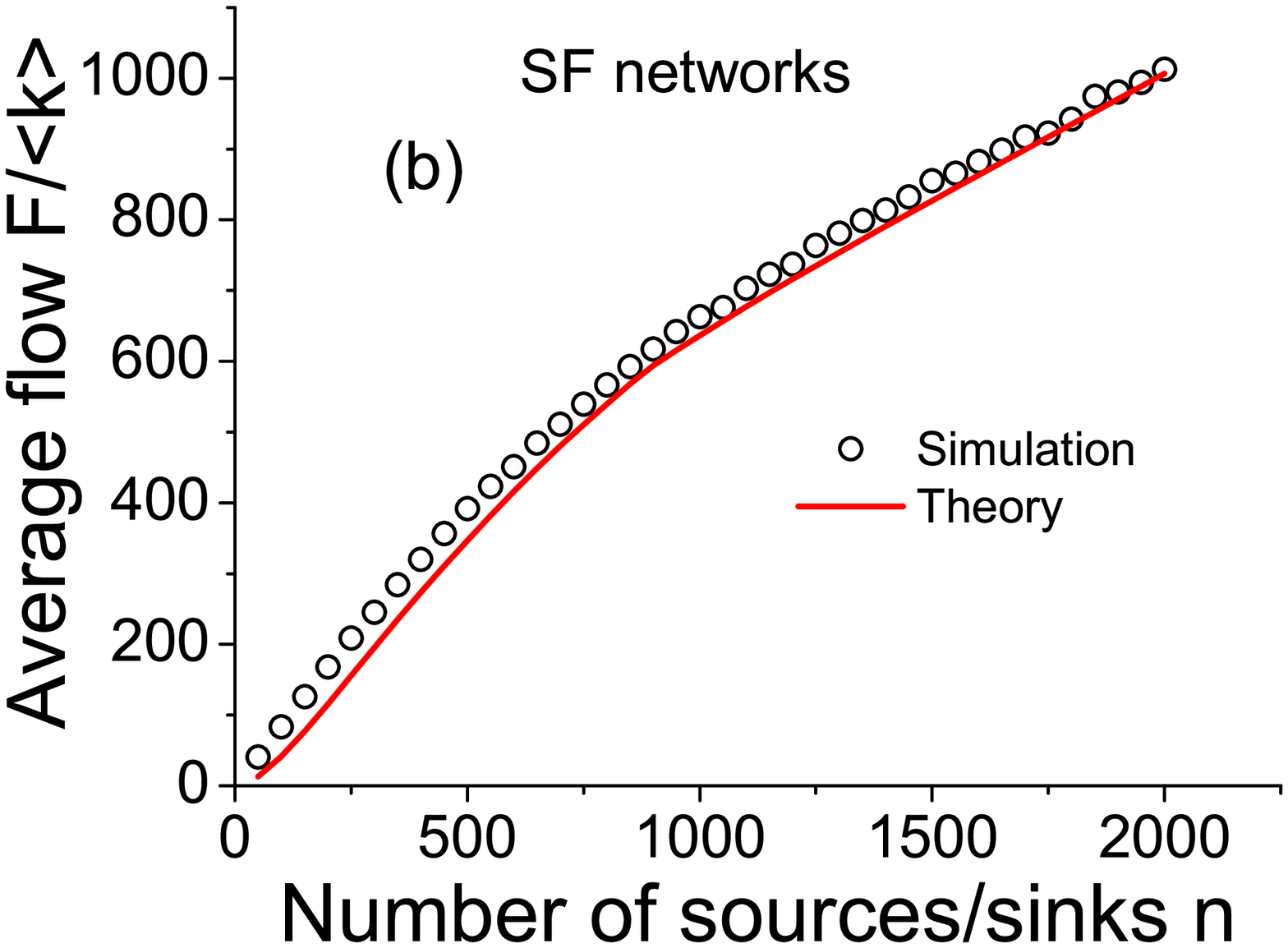}
\end{center}
\caption{Flow in SF networks. (a) Probability distribution of the
flow $\Phi_{n}(F)$ vs. $F$, for $n=1,3,5,10$ (symbols), for the
network of the Internet as of 2007 \cite{Medusa}, which is
approximately scale-free with degree exponent $\gamma \approx 2.5$.
Since $F$ is at least $mn$, a normalized power-law takes the form
$\Phi_{n}(F)\sim n^{2\gamma-2}F^{-(2\gamma-1)}$. Thus, we divide the
vertical axis by $n^{2\gamma-2}$ to make all curves collapse.
Theoretical slope for all $n$'s, $-(2\gamma-1)=-4$ is indicated by
the straight line. (b) Average flow $\overline{F}/\av{k}$ vs. the
number of sources/sinks $n$ for random SF networks. Simulation
results ($N=4096$, $\gamma=2.5$,$m=2$), are shown in circles, lines
represent theoretical curves. For the theory we used the upper bound
for paths of lengths up to three (see main text and Appendix).}
\label{DIMES}
\end{figure}

For SF networks, we approximate the sum in
(\ref{transport_equation}) with an integral:
\begin{equation}
\Phi_n(F) \sim F^{-\gamma}\int_{F}^{\infty}F'^{-\gamma}dF' \sim
F^{-(2\gamma-1)}.
\end{equation}
A similar result holds for the current $I$. This is demonstrated in
Fig. \ref{DIMES}(a) for the internet at the Autonomous Systems (AS)
level \cite{Medusa}. For MC flow, the minimum of the degrees of each
pair has a power-law distribution with exponent $(2\gamma-1)$. From
Eq. (\ref{SFsum}), the tail distribution of the sum remains a
power-law with the same exponent, and therefore also in this case
$\Phi_n(F^{\textrm{MC}}) \sim
\left[F^{\textrm{MC}}\right]^{-(2\gamma-1)}$.

To evaluate the transport for the regime $n \gg 1$, we use a
different approach. Let us concentrate on the case of flow in ER
networks. We condition the total flow on the length of the path
connecting a source and a sink. $n^2$ paths of length one (direct
link between a source and a sink) are possible, and each exists with
probability $p \equiv \av{k}/(N-1)$. Denoting by
$\overline{F_{\ell}}$ the average flow that goes through paths of
length $\ell$, we have $\overline{F_1} = n^2p$.

Paths of length two involve one intermediate node. If the
intermediate node $i$ is connected to $n_s(i)$ sources and $n_t(i)$
sinks, the flow it can channel is
$n_{\min}(i)\equiv\min\{n_s(i),n_t(i)\}$. Since each edge exists
with independent probability $p$, $n_s$ and $n_t$ are binomial
variables with parameters $(n,p)$. The probability for $n_{\min}$ to
take the value $m$ is given by
%\begin{widetext}
\begin{equation}
\label{n_min} P_{n_{\min}}(m) =
2\left[P_{b}(m)\sum_{j=m}^{n}P_{b}(j)\right] -
\left[P_{b}(m)\right]^2,
\end{equation}
where $P_{b}(j)={n\choose j}p^j(1-p)^{n-j}$ since $n_s$ and $n_t$
are binomials.
%\end{widetext}
For large $n$ and small $p$ the binomial variable can be
approximated by a Poisson distribution, for which Eq. (\ref{n_min})
reduces to
\begin{equation}
\label{n_min_closed} P_{n_{\min}}(m) = 2 \frac{(np)^{m} e^{-np}}{m!}
\left( \frac{\gamma(m,np)}{\Gamma(m)} - \frac{(np)^{m} e^{-np}}{2m!}
\right).
\end{equation}
The average of $n_{\min}$ is $ \av{{n_{\min}}} =
\sum_{m=0}^{n}mP_{n_{\min}}(m)$. Since there are $(N-2n)$
possibilities of choosing the intermediate node, $\overline{F_2} =
(N-2n)\av{n_{\min}}$. Note that approximating $\overline{F}\approx
\overline{F}_1+\overline{F}_2$ becomes accurate as $n\rightarrow
N/2$, where there are only very few intermediate nodes and thus a
very small probability for a long path.

Paths of length three involve two intermediate nodes and their
average number is more difficult to compute. However, by a mapping
onto a matching problem we are able to provide lower and upper
bounds for $\overline{F_3}$ (see Appendix). In Fig.
\ref{Theory}(a), we plot $\overline{F}(n)$ obtained with the upper
bound, neglecting flows of higher order $F_4,F_5,...$, and find
agreement with the simulations.

An interesting outcome of the above calculation is an immediate
result for the electrical current. Since all links have unit
resistance, a path of length $\ell$ has total resistance $\ell$ so
$\overline{I} = \overline{F_1}/1 + \overline{F_2}/2 +
\overline{F_3}/3 + ...$~. Our simulation results agree with our
theory, as shown in Fig. \ref{Theory}(b). While it is not in general
correct that in SF networks each edge exists with independent
probability $p$, applying the same approach for SF networks results
in a good qualitative agreement with the simulations---but not as
good as for ER networks (Fig. \ref{DIMES}(b)).

Focusing attention on the transport per source/sink,
$\overline{F}/n$ and $\overline{I}/n$, we observe that while the
flow $\overline{F}/n$ decreases with $n$, the current
$\overline{I}/n$ increases (Fig. \ref{per_user}). The decrease in
the case of the flow is intuitively clear. As more paths become
``bottlenecks'', the total number of paths connecting the sources
and sinks decreases. However, when increasing $n$, the paths that
exist have shorter lengths since the probability for direct or
almost direct link between the sources and sinks is higher. While
this effect does not influence the flow, in the case of electrical
current it reduces the resistivity of the paths and increases the
total current \cite{My_flow2}. This is a fundamental difference
between the two forms of transport, which our analysis reveals.

For MC flow, paths cannot become shorter once a pair of a source and
a sink is added, since the transport takes place between specific
pairs \cite{MCFlow_def} (Fig. \ref{mc_flow}(a)). Therefore, we
expect that for some $n^*$, the network will saturate and will not
be able to carry any more flow. In other words, not only
$\overline{F^{\textrm{MC}}}/n$ will decrease, but
$\overline{F^{\textrm{MC}}}$ itself will not grow.

To develop a theory for the average MC flow in ER networks, we look
at the pairs of sources and sinks as if they are added one at a
time. As the pair $n$ is added, we assume the paths connecting the
previous $n-1$ pairs remain unchanged, such that the new pair can
connect only through these network links not already in use. We also
assume the used links are randomly spread over the network and
denote the average effective degree of the unused part as $k_n$.
Under these assumptions, the new pair will be connected, on average,
with a number of paths that is the minimum of two degrees of average
$k_n$ (see above). Thus, the total MC flow can be approximated as
\begin{equation}
\label{MCavg} \overline{F^{\textrm{MC}}}(n) \approx
\sum_{n'=0}^{n-1}\mu(k_{n'}),
\end{equation}
where $\mu(k)=\sum_{j=0}^{N}2j\frac{k^{j} e^{-k}}{j!} \left(
\frac{\gamma(j,k)}{\Gamma(j)} - \frac{k^{j} e^{-k}}{2j!} \right)$ is
the average of the minimum of two degrees drawn from a Poisson
distribution with average $k$, as in Eqs. (\ref{min_ER}) and
(\ref{n_min_closed}).

To complete the derivation, we need to find the average effective
degree $k_n$. This is straightforward to calculate, if we assume
that to optimize the total flow (since we look at \emph{maximum}
flow), the transport between each new pair uses only shortest paths.
Recalling that in ER networks with average degree $\av{k}$ the
average shortest path is of length $\log{N}/\log{\av{k}}$
\cite{Bollobas}, we can write a recursion equation for the evolution
of $k_n$
\begin{equation}
\label{k_evolution} k_{n+1}= k_n -
\mu(k_n)\frac{\log{N}}{N\log{k_n}}, \quad k_0=\av{k}.
\end{equation}

Evaluating (\ref{MCavg}) and (\ref{k_evolution}), we find agreement
with simulations of the exact MC flow \cite{MCFlow_def} (Fig.
\ref{mc_flow}(b)). Note also, that this formula for the MC flow is
an improvement over the result obtained with the small-$n$
assumption, Eq. (\ref{flow_dist}), as shown in the inset of Fig.
\ref{mc_flow}(b).

 For large $n$ such that $k_n$ is small, the new pair is
no longer guaranteed to be connected. To find the value of  $n^*$,
above which additional sources and sinks cannot communicate, we use
the result of percolation theory that the network becomes fragmented
when the average degree decreases below one \cite{Bollobas}. Thus
$n^*$ satisfies $k_{n^*}=1$. Since substituting $k_{n^*}=1$ in Eq.
(\ref{k_evolution}) does not yield a closed form formula, we bound
$n^*$ by assuming the number of paths connecting the $n$th pair
$\mu(k_n)$ satisfies $1 \leq \mu(k_n) \leq k_n$. Eq.
(\ref{k_evolution}) splits into two inequalities
\begin{eqnarray}
k_{n+1} &\geq& k_n - k_n\frac{\log{N}}{N\log{k_n}} \\
k_{n+1} &\leq& k_n - \frac{\log{N}}{N\log{k_n}} \nonumber.
\end{eqnarray}
Approximating $k_{n+1}-k_n \approx \frac{dk_n}{dn}$, setting
$k_{n=0} = \av{k}$ and solving the two differential inequalities, we
obtain
\begin{eqnarray}
\left(\log{k_n}\right)^2 &\geq& \left(\log{\av{k}}\right)^2 - 2\frac{\log{N}}{N}n \\
\nonumber (k_n\log{k_n}-k_n) &\leq& (\av{k}\log{\av{k}}-\av{k})
-\frac{\log{N}}{N}n.
\end{eqnarray}
Substituting $k_{n^*}=1$, the resulting bounds for $n^*$ are
\begin{equation}
\label{bounds} \frac{\log^2{\av{k}}}{2} \leq
n^*\cdot\frac{\log{N}}{N} \leq \av{k}\log{\av{k}}-\av{k}+1.
\end{equation}
From (\ref{bounds}), $n^*={\cal{O}}(N/\log{N})$, i.e., the maximal
number of sources/sinks pairs that can communicate is of the order
of $N/\log N$.

Due to the long computation time of the exact MC flow, only small
system sizes could be considered in the simulations, thereby leading
to a finite size effect that obscures the percolation transition in
a way that the precise point where the network saturates cannot be
observed (Fig. \ref{mc_flow}(b)). The continuous increase in the MC
flow for $n>n^*$ in our simulations is expected since for finite
networks, even when $k_n<1$ some nodes are still connected
\cite{bunde_havlin}. Moreover, the probability of a pair of nodes to
belong to the same small (non-giant) cluster cannot be neglected for
finite systems.

For SF networks, the degree exponent $\gamma$ plays a significant
role. For $2<\gamma<3$, there is no percolation threshold
\cite{attack}, and we expect more sources/sinks to be able to
communicate in comparison to ER networks. For $\gamma>3$, a
percolation threshold exists, and we expect behavior qualitatively
similar to that of ER networks. However because of the limitation on
computation time, we leave it as a conjecture.

In conclusion, we investigate the efficiency of transport in complex
networks when many sources and sinks are communicating
simultaneously. We obtain analytical results for the total
electrical current and flow when the number of sources/sinks is very
small. For a large number of sources/sinks, we derive approximate
expressions for the total transport in ER networks by looking at the
lengths of the paths used for the transport, and identify a
fundamental difference between the behavior of flow and current. For
multi-commodity flow, we calculate the mean value of the flow in ER
networks. We also argue that in scale-free networks more
sources/sinks can communicate because of the lack of a percolation
threshold.

\begin{acknowledgments}
We thank the ONR, the Israel Science Foundation, the Israel Center
for Complexity Science, and the EU project Daphnet for financial
support; and R. Cohen, M. Allalouf and L. Zdeborova for discussions.
S.C. is supported by the Adams Fellowship Program of the Israel
Academy of Sciences and Humanities.
\end{acknowledgments}

\section{Appendix: Average number of paths of length three}

To calculate the average number of paths of length three that
connect the sources and sinks in ER networks, we note that these
paths take the following form:
\begin{equation}
\label{path} \rm{Source} \rightarrow \rm{IntermediateNode1}
\rightarrow \rm{IntermediateNode2} \rightarrow \rm{Sink}.
\end{equation}
Since the pair of intermediate nodes can be connected by at most one
link, the flow through this pair (in that specific direction) is
either one or zero. To find out whether the path (\ref{path}) is
available, we must first classify each intermediate node $i$ based
on its number of links to the sources $n_s(i)$ and to the sinks
$n_t(i)$. Three classes are possible.
\begin{enumerate}
\item $n_s(i)=n_t(i)$. In this case all the links that connect $i$ to the sources
and sinks are used in paths of length two, $F_2=n_s(i)=n_t(i)$, and
$i$ cannot be used for any $F_{\ell}$ with $\ell>2$.
\item $n_s(i)>n_t(i)$. Node $i$ uses $n_t(i)$ links in
$F_2$, and has $n_s(i)-n_t(i)$ links free to use in $F_{\ell}$,
$\ell>2$. Thus, $i$ can serve as an \emph{IntermediateNode1} in the
path (\ref{path}). We call the set of such nodes ${\cal I}_1$, and
denote its size by $|{\cal I}_1|$.
\item $n_t(i)>n_s(i)$. Node $i$ uses $n_s(i)$ links in
$F_2$, and has $n_t(i)-n_s(i)$ links free to use in $F_{\ell}$,
$\ell>2$. Thus, $i$ can serve as an \emph{IntermediateNode2} in the
path (\ref{path}). We call the set of such nodes ${\cal I}_2$, and
denote its size by $|{\cal I}_2|$.
\end{enumerate}
Each node in ${\cal I}_1$ or ${\cal I}_2$ has at least one free link
to use in $F_3$. One has only to connect as many nodes in ${\cal
I}_1$ to as many other nodes in ${\cal I}_2$, such that the number
of connections of a node in ${\cal I}_1$ will not exceed its value
of $n_s-n_t$ and that the number of connections of a node in ${\cal
I}_2$ will not exceed its value of $n_t-n_s$. For a given node in
${\cal I}_1$, denote the number of its \emph{spare links} $n_s-n_t$
as $s_1$, and define $s_2$ similarly for a node in ${\cal I}_2$. The
maximum flow problem is thus reduced to \emph{a generalized
bipartite matching problem}. This is illustrated in Fig.
\ref{schematic}. Due of symmetry, $\av{|{\cal I}_1|} = \av{|{\cal
I}_2|} \equiv \av{|{\cal I}|}$ and $\av{s_1} = \av{s_2} \equiv
\av{s}$. In addition, these quantities can be calculated. The
probability for an intermediate node to have $n_s > n_t$ is:
\begin{eqnarray}
P(n_s>n_t) &=& P_{b}(1)P_{b}(0) + P_{b}(2)[P_{b}(0)+P_{b}(1)] + ... \nonumber \\
&=& \frac{1 - \sum_{i=0}^{n}[P_{b}(i)]^2}{2} = P(n_t>n_s).
\end{eqnarray}
and therefore $\av{|{\cal I}_1|} = (N-2n)P(n_s>n_t) = (N-2n)\left[1
- \sum_{i=0}^{n}\left[P_{b}(i)\right]^2\right]/2 = \av{|{\cal I}|}$.
The average number of spare links $\av{s_1}=\av{s_2}=\av{s}$ is
\begin{eqnarray}
\av{s}&=&\av{s_1} = \sum_{i=1}^{n}i \cdot P\{(n_s-n_t)=i,~\rm{, given }~n_s>n_t\} \nonumber \\
&=& \sum_{i=1}^{n}i \cdot \frac{P\{(n_s-n_t)=i\}}{P(n_s>n_t)} \\
&=& \frac{2}{1 - \sum_{i=0}^{n}[P_{b}(i)]^2} \cdot
\sum_{i=1}^{n}\sum_{j=i}^{n}i\cdot P_b(j)P_b(j-i). \nonumber
\end{eqnarray}

As a first approximation, we assume $\av{s}=1$, to return to a
regular bipartite matching problem. A recent theorem for ER
bipartite networks has proved that a network with minimal degree at
least 2 has a perfect matching with high probability \cite{Frieze}.
Therefore, a lower bound for the matching size will be the 2-core of
the bipartite network (consisting of the nodes in ${\cal I}_1$ and
${\cal I}_2$), which is given by $x\av{|{\cal I}|}$ where $x =
1-e^{-\beta}(1+\beta)$, and $\beta$ is the solution of
$\frac{\beta}{\av{k}}=1-e^{-\beta}$ \cite{HW_book}. Since each such
matching contributes one unit of flow of length three, the total
matching size $x\av{|{\cal I}|}$ is our lower bound for
$\overline{F_3}$. Neglecting flows of higher orders
$\overline{F_4},\overline{F_5},...$, we obtain a lower bound for the
average flow: $\overline{F}=n^2p+(N-2n)\av{n_{\min}}+x\av{|{\cal
I}|}$. This bound becomes exact in the limit of large $n$, where
there are very few intermediate nodes and thus a very small
probability for a long path.

An upper bound for the flow can be obtained by assuming that each
node in the bipartite graph is able to match (on average) the
minimum between its degree (in the bipartite network) and its number
of spare links $s$. This overestimation of the flow of length three
happens to compensate for neglecting flows of longer lengths, and in
most cases agrees with simulations (see Fig. 3).

\bibliography{cwhs}

\begin{thebibliography}{10}
\expandafter\ifx\csname url\endcsname\relax\def\url#1{\texttt{#1}}\fi

\bibitem{DbA_Book}
\Name{ben Avraham D. \and Havlin S.} \Book{Diffusion and reactions in fractals
  and disordered systems} (Cambridge University Press, New York) 2000.

\bibitem{klafter_nature}
\Name{Condamin S., Benichou O., Tejedor V., Voituriez R. \and Klafter J.}
  \REVIEW{Nature }{450}{2007}{77}.

\bibitem{Lopez}
\Name{L\'{o}pez E., Buldyrev S.~V., Havlin S. \and Stanley H.~E.} \REVIEW{Phys.
  Rev. Lett. }{94}{2005}{248701}.

\bibitem{Li}
\Name{Li G., Braunstein L.~A., Buldyrev S.~V., Havlin S. \and Stanley H.~E.}
  \REVIEW{Phys. Rev. E }{75}{2007}{045103}.

\bibitem{lazaros_pnas}
\Name{Gallos L.~K., Song C., Havlin S. \and Makse H.~A.} \REVIEW{Proc. Natl.
  Acad. Sci. USA }{104}{2007}{7746}.

\bibitem{Dani}
\Name{Bollt E.~M. \and ben Avraham D.} \REVIEW{New J. Phys. }{7}{2005}{26}.

\bibitem{Rieger_rw}
\Name{Noh J.~D. \and Rieger H.} \REVIEW{Phys. Rev. Lett. }{92}{2004}{118701}.

\bibitem{Rieger}
\Name{Lee D.-S. \and Rieger H.} \REVIEW{Europhys. Lett. }{73}{2006}{471}.

\bibitem{My_flow2}
\Name{Carmi S., Wu Z., L\'{o}pez E., Havlin S. \and Stanley H.~E.} \REVIEW{Eur.
  Phys. J. B }{57}{2007}{165}.

\bibitem{ER}
\Name{Erd\H{o}s P. \and R\'{e}nyi A.} \REVIEW{Publ. Math. (Debreccen).
  }{6}{1959}{290}.

\bibitem{Bollobas}
\Name{Bollob\'{a}s B.} \Book{Random Graphs} (Academic Press, Orlando) 1985.

\bibitem{BA_review}
\Name{Albert R. \and Barab\'{a}si A.-L.} \REVIEW{Rev. Mod. Phys.
  }{74}{2002}{47}.

\bibitem{DM_book}
\Name{Dorogovtsev S.~N. \and Mendes J. F.~F.} \Book{Evolution of Networks: From
  Biological Nets to the Internet and WWW} (Oxford University Press, Oxford)
  2003.

\bibitem{PV_book}
\Name{Pastor-Satorras R. \and Vespignani A.} \Book{Structure and Evolution of
  the Internet: A Statistical Physics Approach} (Cambridge University Press,
  Cambridge) 2004.

\bibitem{Molloy_Reed}
\Name{Molloy M. \and Reed B.} \REVIEW{Random Struct. Algorithms
  }{6}{1995}{161}.

\bibitem{Medusa}
\Name{Carmi S., Havlin S., Kirkpatrick S., Shavitt Y. \and Shir E.}
  \REVIEW{Proc. Natl. Acad. Sci. USA }{104}{2007}{11150}.

\bibitem{Coreman}
\Name{Cormen T.~H., Leiserson C.~E., Rivest R.~L. \and Stein C.}
  \Book{Introduction to Algorithms} 2nd Edition (MIT press) 2001.

\bibitem{MF_algorithm}
\Name{Cherkassky B.~V.} \REVIEW{Algorithmica }{19}{1997}{390}.

\bibitem{Wu}
\Name{Wu Z., Braunstein L.~A., Havlin S. \and Stanley H.~E.} \REVIEW{Phys. Rev.
  Lett. }{96}{2006}{148702}.

\bibitem{Redner_book}
\Name{Redner S.} \Book{A Guide to First-Passage Processes} (Cambridge
  University Press) 2001.

\bibitem{MCFlow_def}
Formally, the maximum multi-commodity flow problem is defined as follows
  \cite{Coreman}. There are $n$ commodities defined by the set: $(s_i,t_i),
  i=1,...,n$, where $s_i$ and $t_i$ are the source and sink of commodity $i$,
  respectively. The flow of commodity $i$ along the edge $(u,v)$ is $f_i(u,v)$.
  The total flow in an edge is limited by its capacity $\sum_{i=1}^{n}f_i(u,v)
  \leq c(u,v)$ ($c(u,v)=1$ here for all edges). The maximum multi-commodity
  flow maximizes $\sum_{i=1}^{n}\sum_{w}f_i(s_i,w)$, where the second sum is
  over all neighbors $w$ of the source. The MC flow problem can be recast into
  a linear programming problem, and thus can be solved in polynomial time.
  However, in partice, only small instances (up to about hundred nodes) could
  be solved using the GLPK (http://www.gnu.org/software/glpk/) linear
  programming (interior point) solver that we used. Fast approximation
  algorthims exist \cite{Fleishcer}, which guarantee a solution close to the
  optimal (however, we used only the exact solution).

\bibitem{bottleneck}
\Name{Sreenivasan S., Cohen R., Lopez E., Toroczkai Z. \and Stanley H.~E.}
  \REVIEW{Phys. Rev. E }{75}{2007}{036105}.

\bibitem{bunde_havlin}
\Name{Bunde A. \and Havlin S.} (Editors) \Book{Fractals and Disordered Systems}
  2nd Edition (Springer) 1996.

\bibitem{attack}
\Name{Cohen R., Erez K., ben Avraham D. \and Havlin S.} \REVIEW{Phys. Rev.
  Lett. }{85}{2000}{4626}.

\bibitem{Frieze}
\Name{Frieze A.} \REVIEW{Random Structures and Algorithms }{26}{2004}{319}.

\bibitem{HW_book}
\Name{Hartmann A.~K. \and Weigt M.} \Book{Phase Transitions in Combinatorial
  Optimization Problems: Basics, Algorithms and Statistical Mechanics}
  (Wiley-VCH) 2005.

\bibitem{Fleishcer}
\Name{Fleishcer L.~K.} \REVIEW{SIAM J. Discrete Math }{13}{2000}{505}.

\end{thebibliography}
\bibliographystyle{eplbib}

\begin{figure}[h]
\begin{center}
\includegraphics[width=4cm]{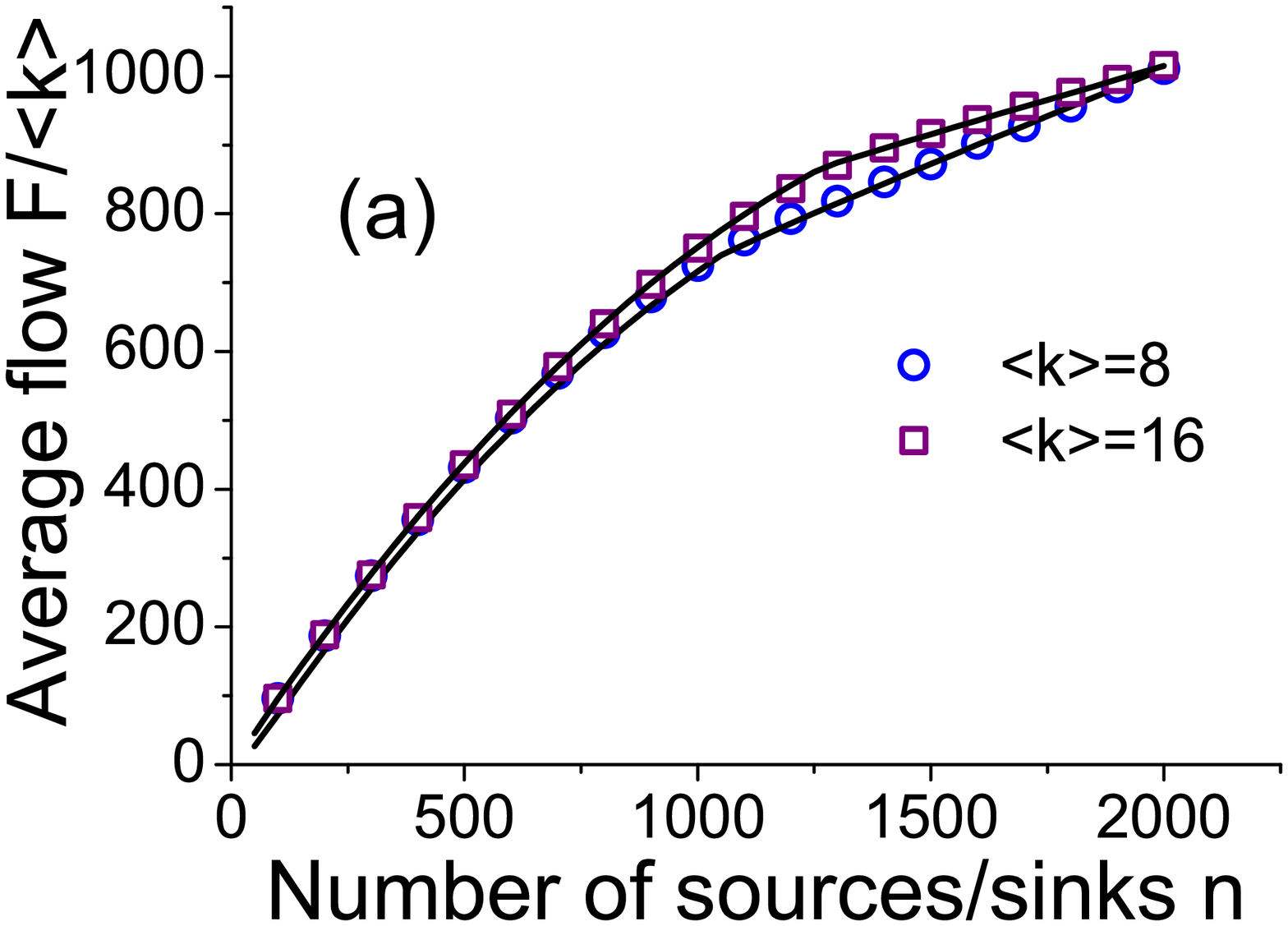}
\vspace{-0.5cm}
\includegraphics[width=4cm]{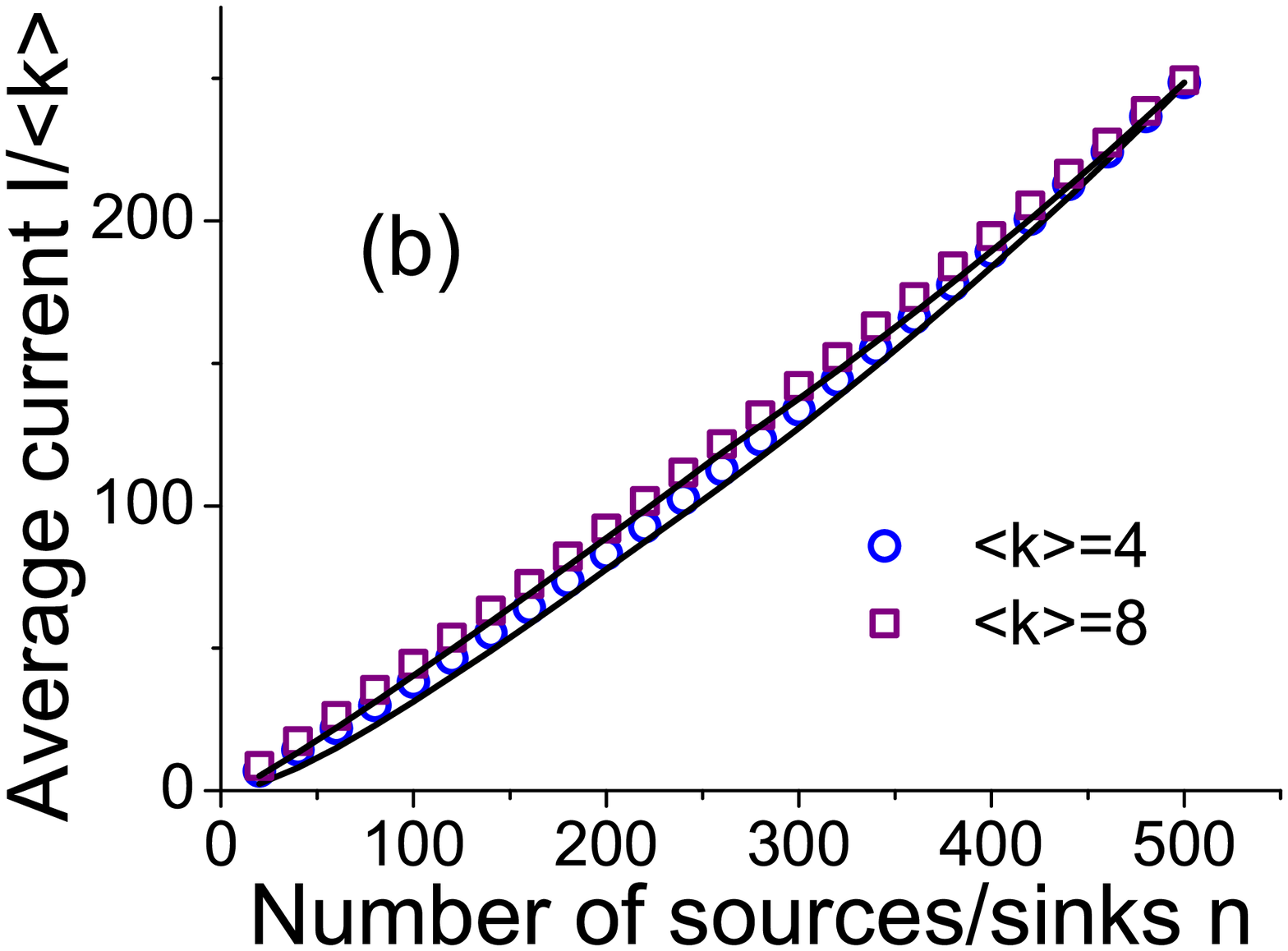}
\end{center}
\caption{Transport in ER networks. (a) Average flow
$\overline{F}/\av{k}$ vs. the number of sources/sinks $n$.
Simulation results ($N=4096$, $\av{k}=8,16$), are shown in symbols,
lines represent theoretical curves. For the theory we used the upper
bound for paths of lengths up to three (see main text and Appendix).
(b) Same as (a) for the average electrical current $\overline{I}$
($N=1024$, $\av{k}=4,8$).} \label{Theory}
\end{figure}

\begin{figure}[h]
\begin{center}
\includegraphics[width=4cm]{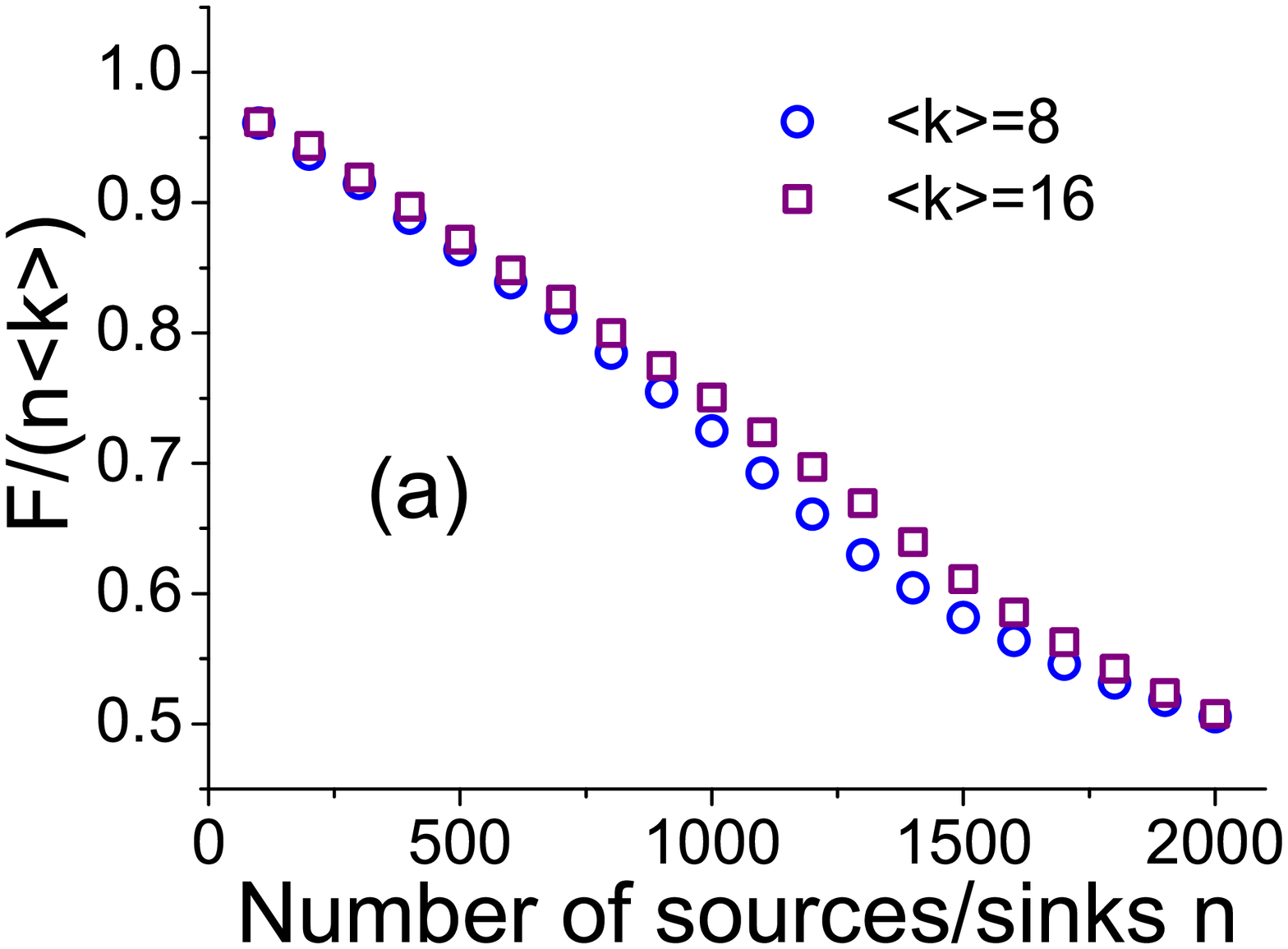}
\includegraphics[width=4cm]{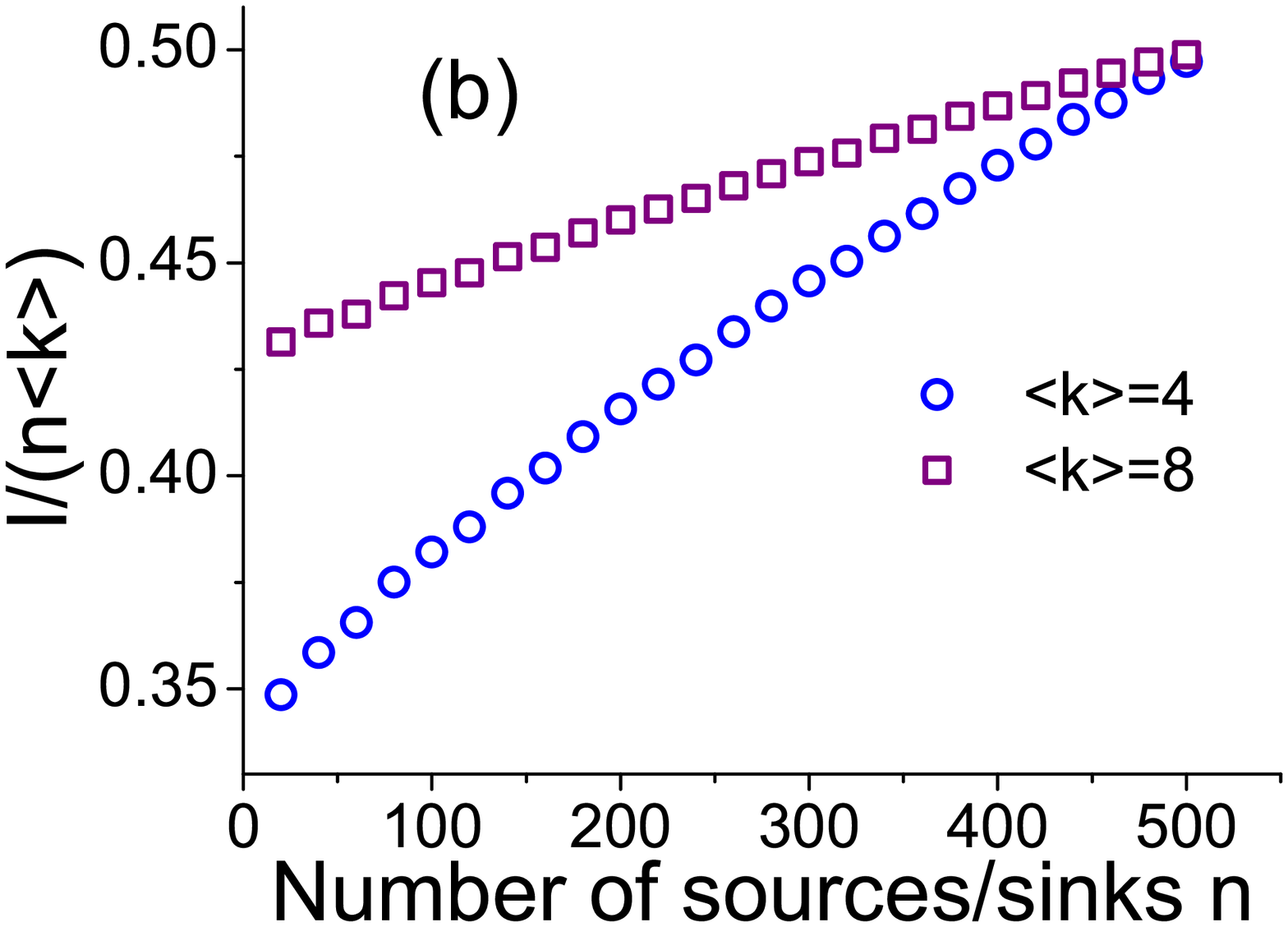} \vspace{-0.5cm}
\end{center}
\caption{Transport normalized by the number of sources/sinks.
Average flow (panel (a)) \emph{per source/sink}
$\overline{F}/(n\av{k})$, and average electrical current per
source/sink $\overline{I}/(n\av{k})$ (panel (b)), vs. the number of
sources/sinks $n$. Note the major difference between the flow and
the current in the behavior of the transport per source/sink: while
$\overline{F}/n$ decreases with $n$, $\overline{I}/n$ increases.}
\label{per_user}
\end{figure}

\begin{figure}[h]
\begin{center}
\includegraphics[width=5cm]{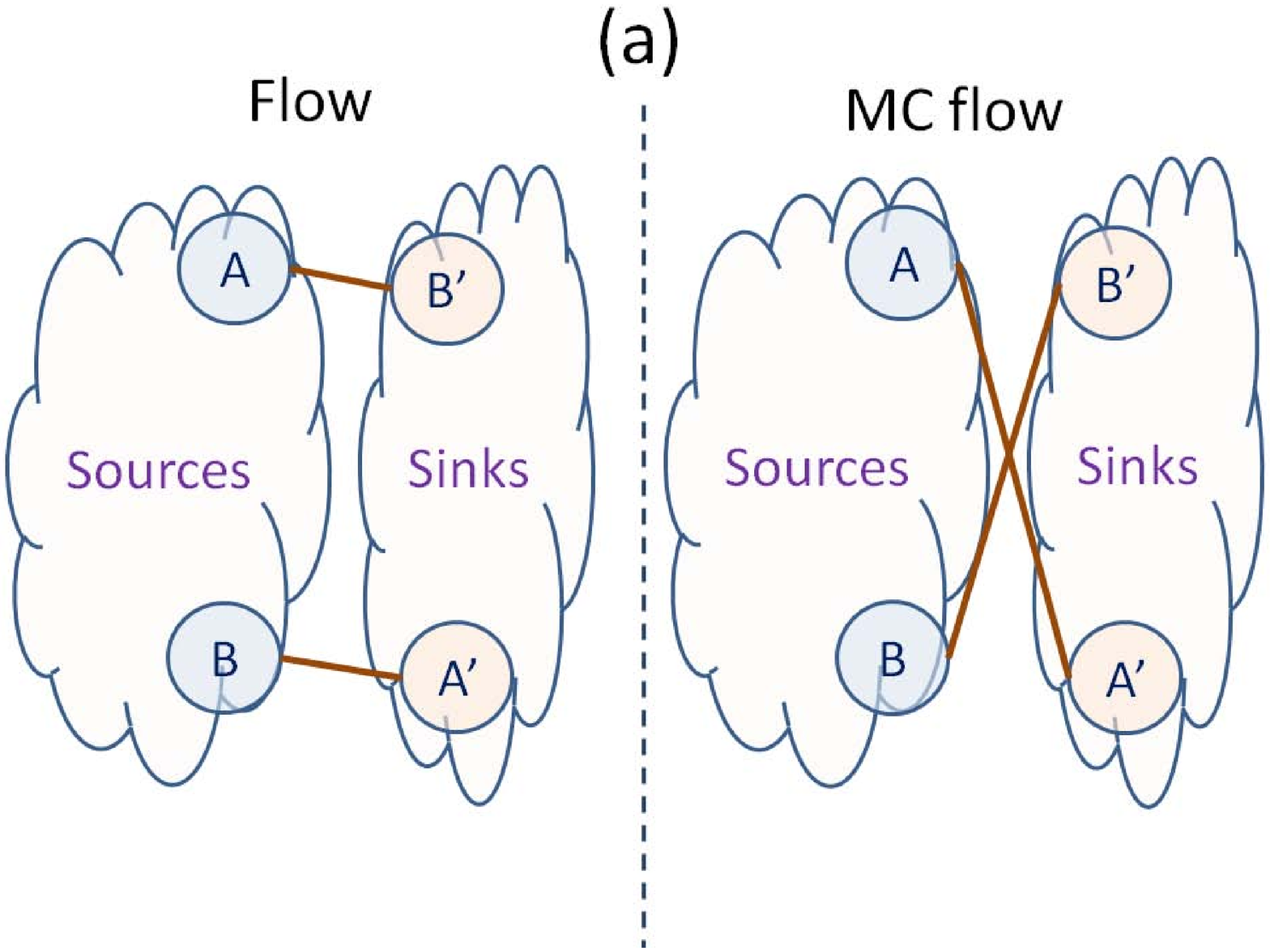}
\includegraphics[width=5cm]{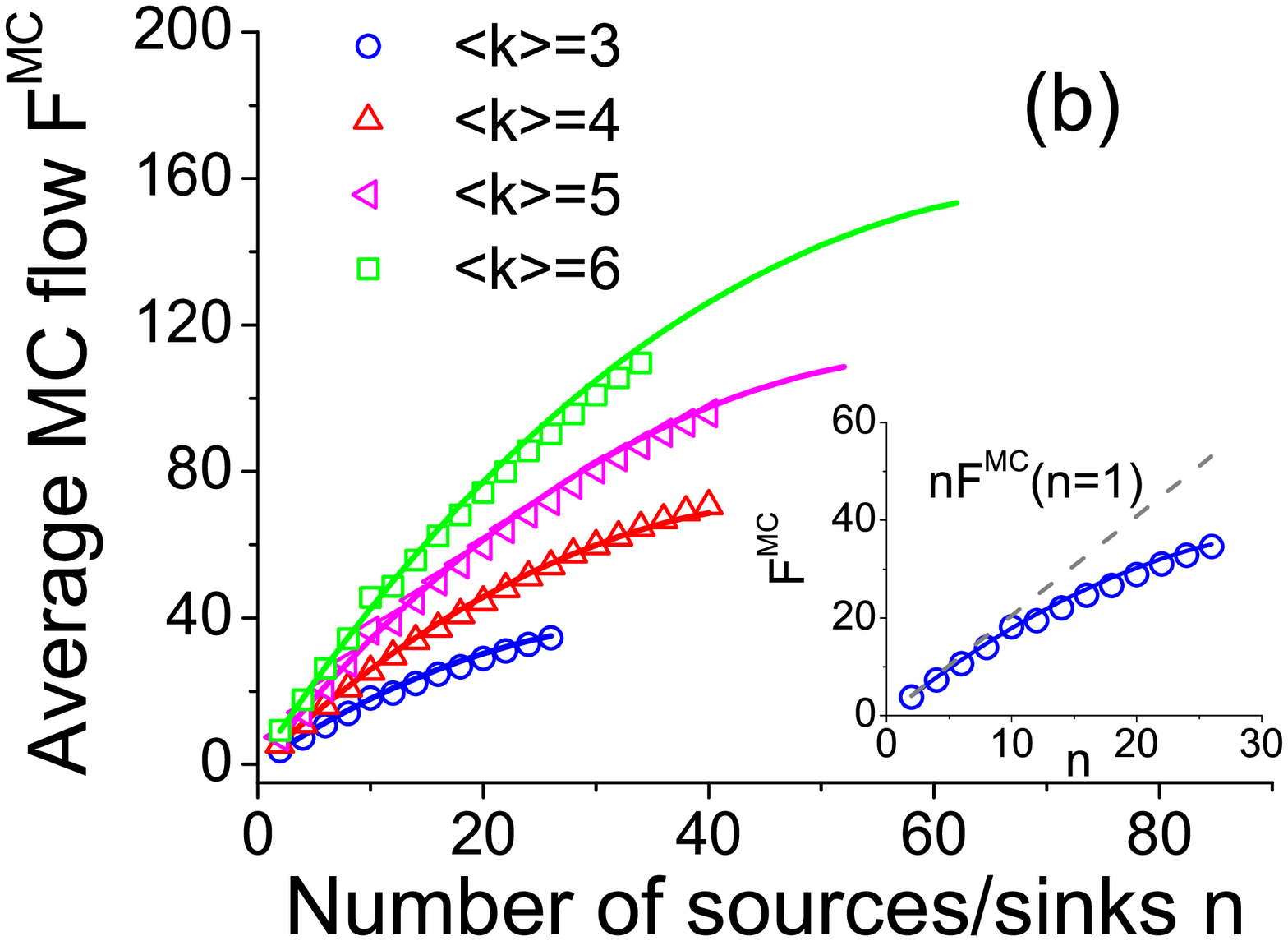} \vspace{-0.5cm}
\end{center}
\caption{Multi-commodity flow in networks. (a) A schematic
illustrating the fundamental difference between flow and MC flow
with respect to path lengths. The sources are $(A,B)$ and the sinks
are $(A',B')$. While for flow any source can connect to any sink,
for MC flow $A$ must connect to $A'$ and $B$ must connect to $B'$,
even at the cost of using longer paths. (b) MC flow
$F^{\textrm{MC}}$ vs. $n$ for ER networks with $N=128$ and
$\av{k}=3,4,5,6$. Symbols correspond to simulations and solid lines
to Eqs. (\ref{MCavg}) and (\ref{k_evolution}) (calculated up to
$n^*$, see text). Inset: for $\av{k}=3$, we compare the simulation
(circles) and theory (Eqs. (\ref{MCavg}) and (\ref{k_evolution}),
solid line) with the small-$n$ approximation
$\overline{F^{\textrm{MC}}}(n)=nF^{\textrm{MC}}_{n=1}$ (dashed
line). Indeed, the small-$n$ approximation overestimates the MC flow
for large $n$, since it does not take into account the decrease in
the average effective degree.} \label{mc_flow}
\end{figure}

\begin{figure}[h]
\begin{center}
\includegraphics[width=6cm]{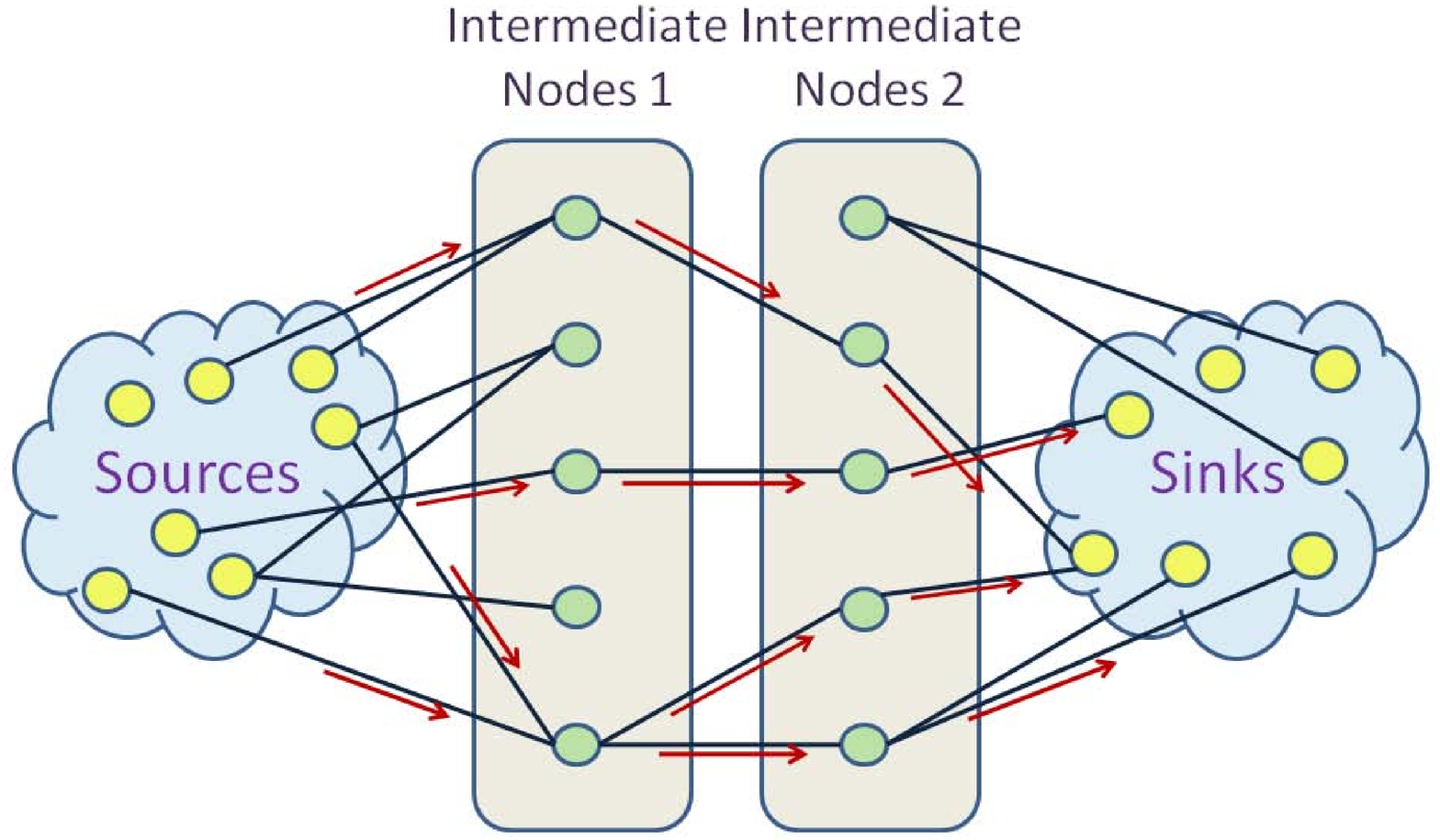} \vspace{-0.5cm}
\end{center}
\caption{A schematic of the bipartite network induced during the
calculation of $F_3$ (Appendix). After taking into account $F_1$ and
$F_2$, we discard all direct links between the sources and the
sinks, as well as all intermediate nodes which have $n_s=n_t$ and
thus all of their links to the sources and sinks are exploited in
$F_2$. This leaves us with two sets: nodes in ${\cal I}_1$ which
have spare links to the sources, and nodes in ${\cal I}_2$ which
have spare links to the sinks. Only these spare links are drawn
here, as well as the links that connect nodes in ${\cal I}_1$ to
nodes in ${\cal I}_2$. Links annotated with an arrow can carry one
unit of flow, in the direction indicated, such that $F_3=4$ in this
example. Note that one intermediate node can channel more than one
unit of flow, by being matched with more than one intermediate node
from the other set.} \label{schematic}
\end{figure}

\end{document}